\documentclass[
twocolumn,
]{ceurart}

\usepackage{listings}
\begin{document}

\copyrightyear{2023}
\copyrightclause{Copyright for this paper by its authors.
  Use permitted under Creative Commons License Attribution 4.0
  International (CC BY 4.0).}

\conference{IEEE International Conference on Advanced Learning Technologies (ICALT23)}

\title{Building Contextual Knowledge Graphs for Personalized Learning Recommendations using Text Mining and Semantic Graph Completion}


\author[1]{Hasan Abu-Rasheed}[%
orcid=0000-0002-2921-4809,
email=hasan.abu.rasheed@uni-siegen.de
]
\cormark[1]
\address[1]{University of Siegen, Siegen, Germany}
\address[2]{Leibniz University Hannover, Hannover, Germany}

\author[1]{Mareike Dornhöfer}[
]

\author[1]{Christian Weber}[
]

\author[2]{Gábor Kismihók}[
]

\author[1]{Ulrike Buchmann}[
]

\author[1]{Madjid Fathi}[
]

\cortext[1]{Corresponding author.}

\begin{abstract}
  Modelling learning objects (LO) within their context enables the learner to advance from a basic, remembering-level, learning objective to a higher-order one, i.e., a level with an application- and analysis objective. While hierarchical data models are commonly used in digital learning platforms, using graph-based models enables representing the context of LOs in those platforms. This leads to a foundation for personalized recommendations of learning paths. In this paper, the transformation of hierarchical data models into knowledge graph (KG) models of LOs using text mining is introduced and evaluated. We utilize custom text mining pipelines to mine semantic relations between elements of an expert-curated hierarchical model. We evaluate the KG structure and relation extraction using graph quality-control metrics and the comparison of algorithmic semantic-similarities to expert-defined ones. The results show that the relations in the KG are semantically comparable to those defined by domain experts, and that the proposed KG improves representing and linking the contexts of LOs through increasing graph communities and betweenness centrality
\end{abstract}


\maketitle

\section{Introduction}

The rapid progress of digitalization in education increased the reachability of learners to online, open educational recourses (OER), e.g., through massive open online courses (MOOCs). While this provides more learning opportunities, an approach of one-size-fits-all on online learning platforms may not work for the wide spectrum of learners, their learning goals, and learning contexts. Personalized learning is proven as an essential requirement for better educational development and learning performance. Moreover, personalization becomes more essential in situations like self-learning, on-the-job training, or vocational education and training (VET), where initial training that learners get, e.g., in formal education, is no longer adequate to the job requirements \cite{buchmann2011, buchmann2022}. This is due to learning taking place within a context, where not only multiple learning objectives may overlap based on the job’s requirements, but also the learner’s background and current state of knowledge might differ considerably. Modern digital learning platforms need to be more contextualized and personalized environments to realize an educational setting that is meaningful to learners, related to their domain, and tailored towards their preferences and needs, to enhance their capacity-development \cite{buchmann2011, buchmann2022}. To account for the learning context, personalization approaches, such as recommender systems (RS), have been developed as context-aware algorithms. Their efficiency, however, is also bound to the contextual information of learning materials that are being recommended, not only to the learner profiles. A personalized, context-aware recommendation is offered when RS algorithms can match a learner’s context to a context of learning objects (LO) and materials. Digital learning platforms still often organize learning content into groups of materials that deploy a hierarchical nature, e.g., as curricula, which are organized into courses, lectures, topics, and finally materials. While this organization is suited to represent the details of a learning objective, it offers no strategy for learning individualization, or model a cross-domain contextualization, which may indicate a learning recommendation, e.g., of courses from different curricula that are required in one application area or one learning context. To consider this context, data models need to be designed in a flexible and semantically enhanced way, which can be realized as a knowledge graph (KG) structure \cite{wilcke2017}. Unlike hierarchical data structures, KG structures offer the potential to consider relations among LOs. Transferring a hierarchical data model into a KG is possible by identifying relations between elements, based on their semantic relatedness and independent of hierarchical limitations. This identification can be done by text mining methods. In KGs, LOs are defined as nodes, and the relations among them are defined as edges. Relations in the KG define one or more types of relevancies between LOs, thus creating a contextual semantic relation based on e.g., the domain of the LOs or their textual similarity. In this paper, we extract information about the shared context of LOs by analyzing their textual semantics. Contextual information is then integrated into the KG as additional relations. Groups of contextually related LOs in the KG may be utilized by a RS, to generate contextual recommendations \cite{trumpower2014, abu-rasheed2021}. We construct the KG utilizing the relations of a hierarchical data model and the semantic relations between LOs, representing the context of those objects in the graph. Semantic relations are extracted with a customized text mining pipeline (TMP), which is designed to analyze and capture the similarities between LOs based on their textual descriptions. A language-dependent similarity algorithm is utilized to mine the semantic similarities between the two description languages of the input data, namely English and German. Our contributions in this paper are: 1) Designing a semantic approach for transferring hierarchical data models in digital learning platforms to graph models. 2) Developing an approach for capturing and modelling the context of LOs to realize semantic KG completion based on a semantic TMP. With the later contribution, present hierarchical structures for LOs can be transformed into contextualizing KGs. In the following sections, an overview of the research background is discussed before we describe the KG construction and semantic relation extraction (RE) among LOs. Then, the evaluation process and results are presented before concluding the paper.

\section{Background}
Historically, learning is founded in “didactic- and lecture-based methods” which oftentimes are focused only on memorizing content instead of supporting “learning transfer” and engendering “problem-solving” \cite{giabbanelli2019}. Wilcke, Bloem and de Boer identified a lack of general data models allowing the representation of more than one domain simultaneously and are thus unsuitable to incorporate heterogeneous data, therefore proposing KGs as a default data model \cite{wilcke2017}. KGs have been utilized in technology-enhanced learning (TEL) \cite{educor_2021, verbert2012, visvizi2019} in recent years. However, the role of the data model in representing the context of LOs in TEL is still less visited in literature, in comparison to e.g. user profiling. Here, we define the context based on situative and subject-oriented learning theories \cite{buchmann2022}, where context is defined as the situation in which learning happens. This includes the learner’s situation (location, previous knowledge, etc.) and the LO’s situation (type, length, implementation scenario, etc.). Context is described by factors that can relate two or more LOs to each other if they serve the same learning situation \cite{verbert2012, hemmler2022}. Ontologies and KGs have been identified as effective methods for embedding LO context into knowledge representation \cite{educor_2021, berri2006}. Furthermore, research shows that the embedding approaches of contextual information in the KG model and algorithms enable capturing the complex relations that learners follow to solve problems \cite{giabbanelli2019, giabbanelli2022, kim2021}. Several approaches have been followed to construct graph models from LOs \cite{fettach2022}. Based on a job-skill hierarchy, authors in \cite{decorte2021, luo2019} utilize the job titles as textual content to create relations between job representations in the KG. To link jobs and their corresponding skills, Dave et al. \cite{dave2018} create three graphs, connecting jobs and skills, using the textual content and existing hierarchical levels. Similarly, de Groot et al. \cite{degroot2021} create a text-based skill matching, with the support of the ESCO framework. The previous approaches, however, do not explicitly address the role of LOs’ context in the KG and do not go far enough to mine further semantic and contextualizing relations between LOs.

\section{Contextual Knowledge Graph Construction}
We develop a KG as a contextual model that represents the semantic relations among LOs. We adopt the taxonomical structure of the digital learning system in \cite{tavakoli2022}, eDoer. This 5-level taxonomy is sufficiently simple and semantically distinctive for the construction of a solid graph data model. Its levels are 1) Journey: which represents a learning goal, 2) Course: a container that represents one aspect of the learning goal, 3) Topic: a concrete knowledge element of a course, 4) Educational Package: a group of related and ordered OERs, and 5) Educational Content: a concrete OER that can be studied. Semantic relations in the KG connect LOs within one or more taxonomical levels. Hereafter, we use the term LO to represent all levels except the Journey. The resulting KG enables modelling the context of a LO through its connections to other objects, which appear in the same learning scenario.

\subsection{Text Mining for Semantic Relation Extraction}
When creating new Journeys and their content, content-creators provide detailed textual descriptions of each object. These descriptions offer the potential to mine semantic similarities between the objects \cite{abu-rasheed2021}. Capturing textual semantics requires text-mining algorithms to consider word meanings and sentence structures. We designed a customized TPM, shown in Fig.1, to maximize the use of LO’s textual metadata. The TMP handles objects’ titles (short texts) and descriptions (long texts), written in two languages (English and German). To detect the language, we use a pre-trained model for language detection \cite{nakatani2010}, with a precision of 99\% over 53 languages. A text-cleaning step then removes special characters and strings that can affect the text similarity algorithm. The structure of the sentences is kept intact, since text similarities are calculated on a semantic level, not as a bag-of-words (BOW). Titles and descriptions follow two processing paths in the TMP, due to the higher volume of description texts, which may include multiple topics. Such topics provide information about potential relevancies to other LOs. Therefore, the next step for titles is creating title embeddings, while descriptions have an additional step of extracting the integrated topics before the embeddings are created for each extracted topic. To generate title embeddings that are accurate in English and German, we use the Sentence-BERT model \cite{reimers2019} and the Spacy library \cite{spacy2023} for natural language processing (NLP) in Python. Title embeddings are used to calculate semantic textual similarities between titles using a cosine similarity algorithm \cite{giabbanelli2019}. Titles with different languages have embeddings created in different corpus spaces. Therefore, we translate German titles before calculating the cosine similarity. The translation is done automatically using DeepL API translation service \cite{deepl2023}. After extracting the main topics from the description using the KeyBERT model \cite{grootendorst2021}, text embeddings are created. To compare two descriptions, an intersection matrix is created to determine which topic-pairs from the two topic sets are compared. The final similarity average is then calculated across all topics of both LOs. The resulting scores from titles and descriptions are used to determine if a semantic relation is created between the two LOs, based on similarity thresholds, which are defined experimentally and then fine-tuned through expert validation.

\begin{figure}
  \centering
  \includegraphics[width=\linewidth]{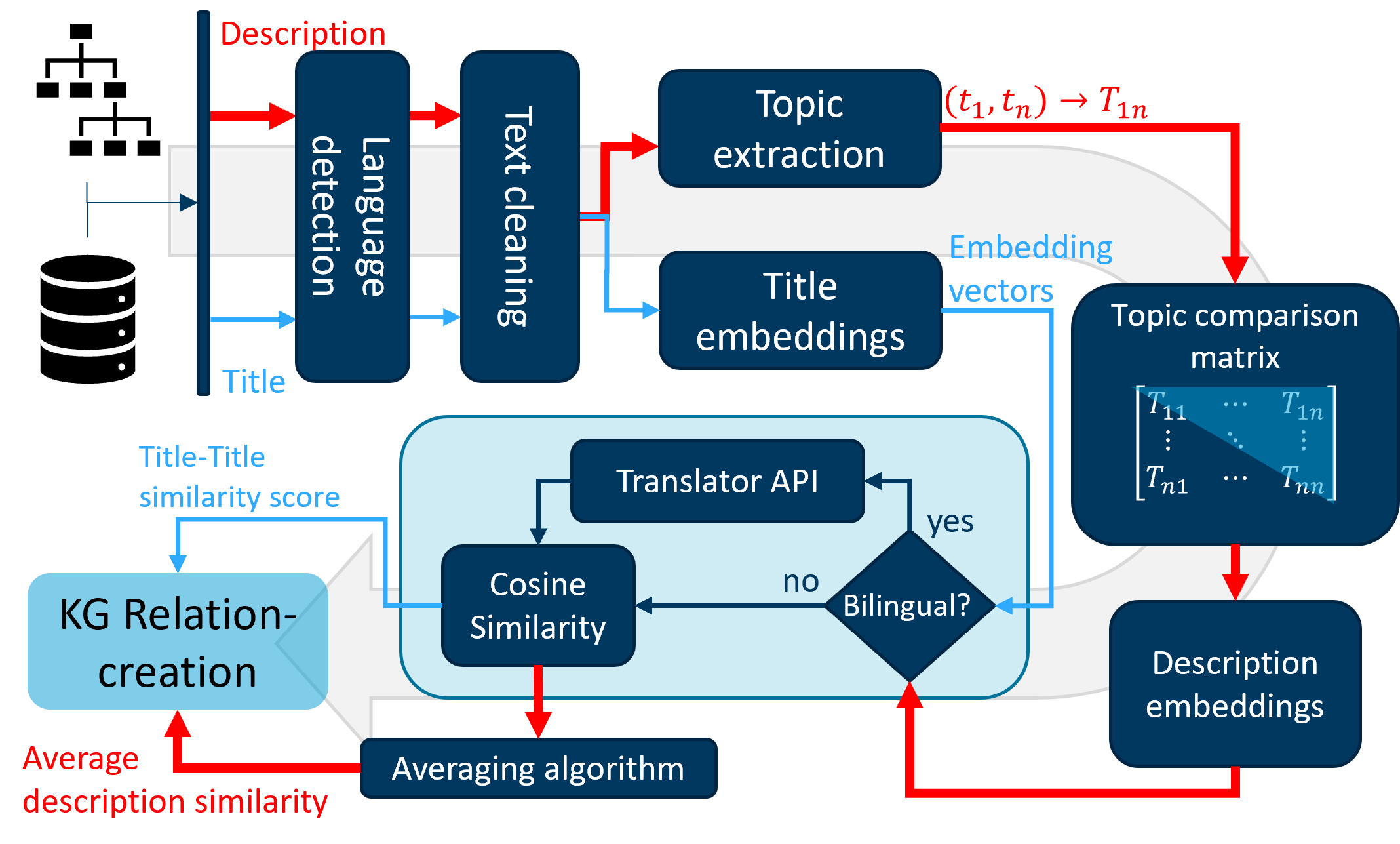}
  \caption{Text mining pipeline (TMP) and semantic similarity calculation for title and description properties of the learning objects.}
\end{figure}

\subsection{Knowledge Graph Construction}
The structure of our KG includes the original hierarchical model and extends it with semantic relations calculated by the TMP, see Fig.2. We preserve the hierarchical relations since they are created by the content creators, and we add the “has\_semantic\_relation\_to” type to them. In the KG, node types represent taxonomy levels. The new relation type creates direct and indirect connections between learning goals (Journeys) and the LOs. For example, although the Journey “Dementia in elderly care” is not directly connected with the Journey “Communication in elderly care", several indirect paths were semantically found through connected Courses and Topics that appear in a similar learning context. This context appears in Fig.2 in the form of a densely connected partial network of Topics (area A), whose textual descriptions revealed that they serve the same scenario. As discussed in \cite{trumpower2014}, such a network can enable a learner to achieve a higher knowledge level that enables comprehension, inferencing, and problem-solving. This is accomplished by learning LOs within their context, in contrast to learning isolated LOs that may only enable remembering.

\begin{figure}
  \centering
  \includegraphics[width=\linewidth]{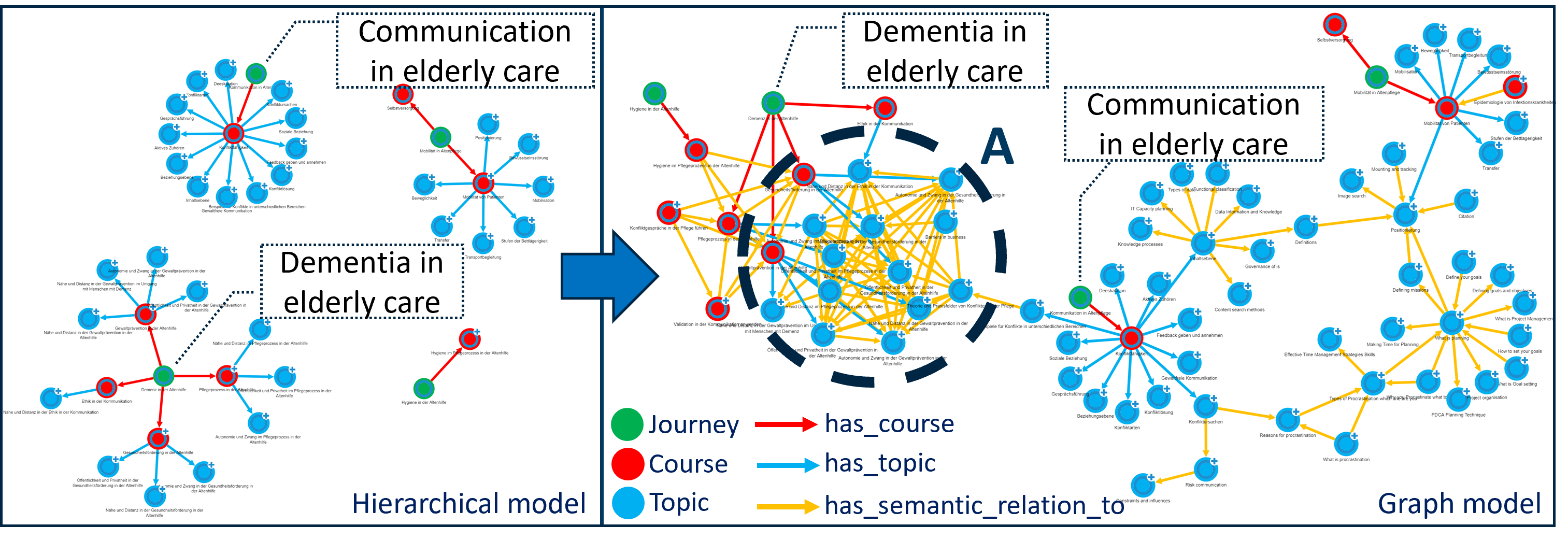}
  \caption{First three levels of the KG structure, representing the role of semantic relations in creating the KG from the hierarchical one.}
\end{figure}

\section{Evaluation and Results}

The evaluation of KG’s quality utilizes a wide range of metrics, which may have different indications in different domains. In TEL and VET, it is important to consider the metrics, whose indications can reflect the role of the KG in the learning process. KG evaluation should also address the quality of the extracted relations, as a foundation for evaluating the quality of the KG structure. Therefore, our evaluation strategy is implemented quantitatively and qualitatively. Qualitative evaluation is conducted with domain experts to validate the relation-extraction process and results, as well as the KG’s role in contextualizing the learning recommendations. Quantitative evaluation is designed two-fold: The first part evaluates the quality of extracted relations in the KG. It focuses on the TMP and evaluates its ability to connect LOs semantically in comparison to expert-created links. The second part considers network metrics \cite{gueret2012} and quality-control methods, as surveyed in \cite{zaveri2016}, where those are classified in the domain of linked open data (LOD) into six dimensions: representation, accessibility, intrinsic, dataset dynamicity, contextual, and trust. To select the metrics that are relevant and meaningful in the scope of this research, we utilize \cite{giabbanelli2019, shin2021, kim2021}. They show that effective evaluation of KG’s ability to represent learning context and the learner’s mental model can be accomplished through metrics that correspond to contextual dimension (relevancy and completeness) and intrinsic dimension (interlinking, density, average degree). We build on those findings and use the metrics of \textit{average degree centrality}, \textit{clustering coefficient}, \textit{weakly connected components}, and \textit{betweenness centrality}.

\subsection{Evaluation Dataset}
For the evaluation, we use a dataset of expert-curated OERs. The data set is provided by Tavakoli et al. \cite{tavakoli2022}. Expert-curated content provides a high level of credibility to the baseline model, which is essential for a fair evaluation. After data exploratory analysis (DEA), we filter out LOs that do not have educational content associated with them and remove the duplicated and isolated ones. The resulting hierarchical model we use for creating the KG features 122 Journeys, 432 Courses, 767 Topics, 2565 Educational packages, and 7358 OERs.

\subsection{Evaluation of the Semantic Relation Extraction}

The goal of semantic relation extraction is to connect LOs that appear in the same or a similar learning context. This goal aligns with the best practices of content creators, who connect an LO to another one if both are needed in the same learning situation. We utilize this logic to evaluate the semantic similarity scores that our TMP calculates among Courses and Topics. We first calculate the average of text cosine similarity scores among the LOs in each Journey (i): 
\begin{math}
    J_{i}^{sim}
\end{math}, 
by comparing each LO to all other LOs in Journey i. Then, we compare the similarity score of each semantic relation \begin{math}
    SR_{i,j}^{sim}
\end{math}, 
to the average similarity of the journeys i,j that it connects \begin{math}
    J_{i,j}^{avg_{sim}}=\frac{J_{i}^{sim}+J_{j}^{sim}}{2}
\end{math}. 

If the TMP is able to extract semantic relations with similarity scores that are comparable to the expert-curated relations, we assume that the new semantic relation is meaningful. We define comparability here as being within the same range of both Journeys’ similarity \begin{math}
    J^{sim}
\end{math}. This means that \begin{math}
    SR_{i,j}^{sim}>=J_{i,j}^{avg_{sim}}
\end{math}. Fig.3 shows the scores of a random sample of 240 semantic relations in the KG, alongside the semantic similarity of each Journey-pair they connect. From the results, we can conclude: 1) the similarity scores calculated within each Journey range between 86\% and around 90\%. This confirms the basic assumption that expert-curated LOs in Journeys are semantically similar. 2) 79\% of our semantic similarity scores are either equal or higher than score averages calculated within Journeys, while the remaining 21\% is slightly below the average. This indicates that the semantic relations between different Journeys are as meaningful as the relations that experts created within each Journey.

\begin{figure}
  \centering
  \includegraphics[width=\linewidth]{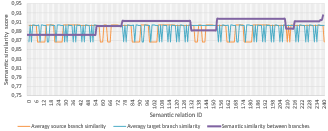}
  \caption{Semantic-relation textual-similarity scores (Purple), compared to the Journeys (Blue and Orange) connected by those relations through the TMP.}
\end{figure}

\subsection{KG Evaluation Metrics}
From the wide range of graph quality metrics, our selection is based on the meaning of each metric in the scope of TEL and VET. Selected metrics reflect certain aspects of our KG, when interpreted from a pedagogical point of view in TEL and VET. The comparison here is conducted between the hierarchical structure and the proposed KG, in the light of the value-added that KG structure offers based on the interpretation of each graph quality metric. In the following, we elaborate on the selected metrics and their domain interpretation. Table 1 shows our KG results in comparison to the original hierarchical model.

\subsubsection{Average Degree Centrality (ADC)}
Degree centrality (DC) measures a node’s popularity in a graph \cite{freeman1978centrality}. It evaluates the connectedness of a node through its incoming and outgoing relations. In an educational use-case, ADC reflects LO’s connectedness to other objects that take place in the learning scenario. The reason is that LOs are either connected by experts, or through a semantic relation. While DC in a hierarchical model represents the LO’s relation to other LOs from the same Journey, the increase of that object’s DC in the graph model shows that new relations have been found to other Journeys, meaning that the LO may appear in more learning contexts. After calculating the DC for all graph nodes, the average (ADC) is determined, which has doubled for our KG.

\subsubsection{Clustering Coefficient (CC)}
CC reflects how strongly a node belongs to a community in a graph. Communities are clusters of nodes that are densely inter-connected. A community is considered more modular if it is easy to separate from other communities. Communities of LOs in our KG represent their context. Therefore, the higher the community count is, the more contexts are represented. Modularity, on the other hand, reflects the potential to separate a learning context from another. Therefore, a lower modularity score is preferred, since it means that learning contexts are well connected to each other. Evaluation results show that our KG increased the detected communities, due to the semantic relations, and decreased graph modularity.

\subsubsection{Weakly Connected Components (WCC)}
WCC detects sets of nodes in the graph that are loosely connected to other graph parts. Therefore, a high WCC score reflects poor connectedness of LO groups in the data model. Using Monge and Elkan's algorithm \cite{monge1997efficient}, we find an effective reduction of the loosely connected node groups in our KG.

\subsubsection{Betweenness Centrality (BC)}
BC describes the impact of a node on the flow of information within that graph. It is especially relevant to VET and TEL, since LOs with high BC work as bridges that link individual learning goals to each other, to solve a more complex problem \cite{trumpower2014}. Using Brandes and Pich's algorithm \cite{brandes2007centrality}, our results show that the KG has an average BC value of 15.1, which is about 10 times the BC score of the original hierarchical model.

\begin{table*}
  \caption{KG Quality Evaluation Against Hierarchical Model. Preferred Value-Trends in VET and TEL Use-Case are Explained }
  \begin{tabular}{lccc}
    \toprule
    Evaluation Metric&Hierarchical data model&KG&Preferred value trends\\
    \midrule
    Average Degree Centrality	&1.079	&2.262	&increasing\\
    Clustering Coefficient
(Number of communities)	&253	&541	&increasing\\
    Clustering Coefficient
(Average modularity score)	&0.779	&0.636	&decreasing\\
    Weakly Connected
Components	&63	&35	&decreasing\\
Betweenness Centrality	&1.57	&15.1	&increasing\\
  \bottomrule
\end{tabular}
\end{table*}

\subsection{Qualitative expert-evaluation}
We also evaluated our KG construction and contextualization approach with domain experts in two focus groups. The groups included experts in VET and researcher-training programs. A total of nine experts participated and answered three questions on 1) The role of connecting learning goals in learning contexts, 2) KGs as contextual data models in their domain, and 3) Using LO’s textual description to mine contextual relations. All participants emphasized the need for connecting individual learning goals for solving real-life, job-related problems. Six participants pointed out that highly connected nodes in the graph can represent transferrable skills among different domains. Three experts found a direct use of connecting multi-lingual LOs in their daily work. They also pointed out that a recommendation of multi-lingual content should avoid repetition of the same content written in different languages. Experts also agreed that textual content and description of LOs are important sources for contextual information. They indicated that content creators should roughly understand the way algorithms extract semantic relations, so that they can enrich the descriptions of their LO content with useful contextual information, allowing better connectivity in the KG. The pedagogical experts also addressed the dynamic change of the learner’s context, and the development of the learning domain, which requires a continuous update of the KG.

\section{Conclusion}
In this paper, we introduced a semantic approach for KG completion to enhance the contextual representation of LOs for personalized learning systems. A concept and a text-mining pipeline for relation extraction are proposed, to transfer hierarchical data models into graph ones, thus enhancing the structural and contextual quality of the data model. Our findings from the TMP and KG evaluation suggest that the KG was able to enhance LOs connectivity on a semantic level. Increased connectivity allowed the KG to represent a context around an LO, through its relations to other similar LOs, which appear in the same or similar learning contexts. Proposed TMP can be used with different hierarchical structures of LOs, such as those generated from other digital learning platforms, since it utilizes the commonly used title- and description properties of LOs. A limitation here is the dependency of our TMP on the volume and quality of LO’s textual metadata. The multilingualism of the proposed solution corresponded to a real-world challenge in VET, but it also raised the concern about repetitive content in a KG-based RS. Although our solution is not responsible for the decision-making process in that scenario, it can still be further developed to expand the similarity description, reflecting potential identical content in multiple languages. Further steps of this research aim to increase the robustness of TMP against textual data sparsity and enrich LOs contextualization with additional domain-specific features.

\bibliography{sample-2col}

\end{document}